\documentclass[manuscript,authorversion,nonacm]{acmart}

\usepackage{amsmath,amsfonts,amsthm}
\usepackage{algorithm,algorithmic}
\usepackage{xspace}
\usepackage{booktabs}
\usepackage{graphicx}
\usepackage{textcomp}
\usepackage{xcolor}
\usepackage{listings}
\usepackage{multirow}
\usepackage{relsize}
\usepackage{pifont}
\usepackage{hyperref}
\usepackage{subfig,paralist}
\usepackage{pifont}
\usepackage{enumitem}
\usepackage{url}
\usepackage{booktabs}
\usepackage{tabularx}
\usepackage{multirow,array}
\usepackage{threeparttable}
\usepackage{bm}

\usepackage{tikz}

\def\BibTeX{{\rm B\kern-.05em{\sc i\kern-.025em b}\kern-.08em
    T\kern-.1667em\lower.7ex\hbox{E}\kern-.125emX}}

\newtheorem*{theorem*}{Theorem}

\usepackage{tcolorbox}
\definecolor{mycolor}{rgb}{0.122, 0.435, 0.698}
\definecolor{gray1}{gray}{0.3}

\definecolor{codegreen}{rgb}{0,0.6,0}
\definecolor{codegray}{rgb}{0.5,0.5,0.5}
\definecolor{codepurple}{rgb}{0.58,0,0.82}
\definecolor{backcolour}{rgb}{0.95,0.95,0.92}
\lstdefinestyle{mystyle}{
    commentstyle=\color{codegreen},
    keywordstyle=\color{magenta},
    numberstyle=\tiny\color{codegray},
    stringstyle=\color{codepurple},
    basicstyle=\tiny\ttfamily,
    breakatwhitespace=false,
    breaklines=true,
    captionpos=b,
    keepspaces=true,
    numbers=left,
    numbersep=5pt,
    showspaces=false,
    showstringspaces=false,
    showtabs=false,
    tabsize=2,
    columns=fixed
}
\definecolor{darkgreen}{rgb}{0.0, 0.5, 0.0}
\definecolor{darkred}{rgb}{0.82, 0.1, 0.26}
\begin{document}
\title{Automatic Programming: Large Language Models and Beyond}

\author[1]{Michael R. Lyu}
\email{lyu@cse.cuhk.edu.hk}
\affiliation{\ Chinese University of Hong Kong, Hong Kong \city{Hong Kong}\country{China}}

\author[2]{Baishakhi Ray}
\email{rayb@cs.columbia.edu}
\affiliation{\ Columbia University\city{New York}\country{USA}}

\author[3]{Abhik Roychoudhury}
\email{abhik@comp.nus.edu.sg}
\affiliation{\ National University of Singapore,  {\em [Corresponding Author]}\city{Singapore}\country{Singapore}}

\author[4]{Shin Hwei Tan}
\email{shinhwei.tan@concordia.ca}
\affiliation{\ Concordia University \city{Montreal}\country{Canada}}

\author[5]{Patanamon Thongtanunam}
\email{patanamon.t@unimelb.edu.au}
\affiliation{\ University of Melbourne\city{Melbourne}\country{Australia}}

\begin{abstract}
Automatic programming has seen increasing popularity due to the emergence of tools like GitHub Copilot which rely on Large Language Models (LLMs). At the same time, automatically generated code faces challenges during deployment due to concerns around quality and trust. In this article, we study automated coding in a general sense and study the concerns around code quality, security and related issues of programmer responsibility. These are key issues for organizations while deciding on the usage of automatically generated code. We discuss how advances in software engineering such as program repair and analysis can enable automatic programming. We conclude with a forward looking view, focusing on the programming environment of the near future, where programmers may need to switch to different roles to fully utilize the power of automatic programming. Automated repair of automatically generated programs from LLMs, can help produce higher assurance code from LLMs, along with evidence of assurance.
\end{abstract}

\maketitle

\renewcommand{\shortauthors}{Authors}

\section{Challenges in Automatic Programming}


The task of programming both in terms of intent capture (capturing the desire of the user) as well as in generation of correct code --- has occupied much of the Computing profession for the last 50--60 years. There has been significant progress in modeling and system design to support accurate intent capture leading to the growth of formal specifications. However, despite all the progress – software engineers are reluctant to write formal specifications, and for large software systems a formal description of intent is not available --- leading to tremendous hardship in debugging and fixing errors. The field of automated program repair has shown promise in terms of code generation at a micro-scale. The key question there is how to trust the automatically generated code. 

Recent developments on automated code generation from Large Language Models (LLMs) bring the trust issues in auto-coding even more into the forefront. This raises not only the overall question of correctness of automatically generated code, but at what point we can start trusting automatically generated code enough to integrate it into our code-base. In past decades, niche industries have generated code from models, however there is no precedent of automatically generated code from natural language specifications being used widely. We discuss the trust issues for such automatically generated code thoroughly in this article.  While the immediate motivation of the article is to study the trust issues in code from Large Language Models (LLMs), we study the topic of automated programming more broadly in this article.

We notice that increasingly many organizations are moving towards automatically generated code, even apart from the popularity of large language models. A recent keynote at Oracle CloudWorld 2023 \cite{oracle} mentions how Oracle is considering moving away from writing software in Java, and instead automatically generating code for new software projects in a language called Apex. Apex is a well-known low code application platform to assemble an application out of application pages. This movement towards low code enables other benefits such as easily achieving security audit of a software project. Overall, we note that automatic programming goes beyond the use of large language models and implicitly includes recent trends in the growth of {\em low-code no-code} application development.

As a result of the recent interest in automatic programming, the set of problems associated with automatically generated code have received wide attention. Apart from correctness, there are concerns about security, privacy, explainability of the code - particularly when generated from a large language model. Pragmatically speaking, there could remain concerns about ``passing the blame'' when a software project which includes automatically generated code fails. To understand the underlying issue, we can draw an analogy between interaction between application software and systems software which leads to the well-known {\em application compatibility} (often called {\tt appcompat} by developers) problem (e.g. see \cite{microsoft}). Typically due to version change of systems software such as operating system (OS), a specific application running on top of the OS (such as PDF reader) may fail. However, this need not be owing to the operating system itself. It may be due to a mistaken understanding of the {\em expectations} between the application software and the OS. In a similar fashion, when automatically generated code and manually written code co-exist in a software project, errors may creep in due to mistaken understanding of the expectations between different software components. 

In this article, we thus examine how trust boundaries can shift when we integrate automatically generated code into a software project. One of the technical questions that could be of interest for the research community are the acceptability criterion for integrating LLM generated code into a software project. The capability of LLMs augmented by program analysis tools in automating key programming tasks such as bug-fixes and feature additions, as articulated in the recently proposed SWEbench \cite{swe} is also worthy of study. The last mile improvement of automatically generated code from LLMs, through the systematic use of program repair \cite{fan2023automated} remains a possibility to explore~\cite{liu2023automated}. 
Such a penultimate auto-repair strategy can help provide specific evidence of correctness (such as passing curated tests) that builds confidence towards accepting LLM-generated code into a code repository. 

We also study the impact of LLMs in automating non-code artifacts and processes such as test generation, code review and code summarisation. More importantly from the human-LLM interaction perspective, we seek to provide an emerging new outlook in future day programming. Traditionally, when formal specifications are unavailable, software engineers have resorted to {\em program comprehension} or specification inference to understand the functionality of a complex software system. This practice is particularly relevant when the software system is not built as a monolithic  artifact, but rather assembled via the co-operation of different teams or via open-source contributions. 

We note that the traditional program comprehension problem is conducted by the human, and may involve the usage of analysis / debugging tools to understand the working of a complex software system. In the age of LLM driven programming, we could postulate a {\em new comprehension problem} - where based on the natural language requirements LLMs augmented by program analysis tools can automate bulk of the comprehension tasks. There could be structured provisions for consulting the human to disambiguate requirements, at different stages of the comprehension process. Studying the mechanisms for such human-LLM collaboration and providing adequate primitives for consulting the human programmer by the LLM / analyzers could point us to the programming environments of the near future. We also underline the possibility of automated program repair of automatically generated code as a flexible mechanism for trusted automatic programming. These could feature in future day programming environments of 2030-35 and beyond. 

\section{Historical milestones and literature review}


\textcolor{black}{In this section, we will delve into the development of automatic programming, pinpointing historical milestones and conducting a thorough literature review. Specifically, we will first introduce several key tasks that propelled the field forward including code generation, program repair, and software testing. Additionally, we will highlight recent advances of LLMs in automatic programming and logging.}

\subsection{Code Generation}
\textcolor{black}{Code generation, also known as program synthesis, refers to the automated generation of software code based on user intent. This technique boosts developer productivity by reducing manual coding and accelerating the software development lifecycle. Early research in code generation mainly centers on deductive and inductive program synthesis which crafts code based on specification and/or input-output pairs. With the advent of deep learning techniques, natural language-based code generation that describes users' intent in plain language gains prominence. Besides, there are also other works that focus on generating code based on images and structured data. In this paper, we will mainly introduce deductive, inductive, and natural language-based code generation methodologies.}

\paragraph*{Deductive and Inductive Program Synthesis}
\textcolor{black}{Deductive program synthesis crafts programs from high-level descriptions, which involves mechanical theorem proving and formal methods~\cite{green1969theorem,gulwani2017program}. The requirement of detailed specifications helps reduce logical errors. It has diverse applications in fields such as robotics~\cite{fikes1971strips} and software engineering~\cite{DBLP:journals/csur/HieronsBBCDDGHKKLSVWZ09}. For example, STRIPS~\cite{fikes1971strips} is an automatic planner that addresses robot problems, and PROW~\cite{DBLP:conf/ijcai/WaldingerL69} generates LISP code from specifications in predicate calculus by incorporating a two-step process involving theorem proving and code production. Another work~\cite{koza1994genetic} uses genetic programming approaches to automatically evolve programs that are consistent with a specification. Conversely, inductive program synthesis, or programming by example (PBE), generates programs directly from specific input-output pairs. This approach is less complex and more user-friendly than deductive synthesis and has been extensively investigated. It enables users unfamiliar with programming to instruct computers through examples. For instance, FlashFill~\cite{DBLP:conf/popl/Gulwani11}, one of the most popular real-world program synthesis application, generates programs for spreadsheets like Excel from very few input–output examples. Similar methods~\cite{DBLP:conf/pldi/0001DSD19} are also used to generate programs for relational databases for schema refactoring.}

\paragraph*{Natural Language-based Code Generation}
\textcolor{black}{Existing natural language-based code generation mainly employs deep-learning techniques and can typically be divided into three categories: sequence-based, tree-based, and pre-trained model approaches. In the realm of sequence-based models, the generation process employs the sequence-to-sequence paradigm and treats this process as a machine translation process to translate the natural language description into source code. For instance, Ling et al.~\cite{DBLP:conf/acl/LingBGHKWS16} employ a neural network with a structured attention mechanism to handle semi-structured inputs for code generation. For tree-based models, these methods take into account the inherent structured nature of programs and parse them into a tree such as an Abstract Syntax Tree (AST). For example, Yin et al.~\cite{DBLP:conf/emnlp/YinN18} train an LSTM to generate a sequence of tree-construction actions and subsequently build the AST from these actions. Rabinovich et al.~\cite{DBLP:conf/acl/RabinovichSK17} propose the Abstract Syntax Networks and directly generate the tree structure of source code. Another work~\cite{DBLP:conf/aaai/SunZXSMZ20} design Transformer blocks to encode both the natural language and the previously generated grammar rules and then predict subsequent grammar rules in the sequence. In recent years, the advent of pre-trained models~\cite{DBLP:conf/emnlp/FengGTDFGS0LJZ20,DBLP:conf/emnlp/0034WJH21} has achieved significant improvement in the field. These models are pre-trained on extensive datasets and then fine-tuned on datasets related to code generation. Furthermore, some studies draw inspiration from code reuse practices to enhance code generation models using retrieval-augment generation. Hayati et al.~\cite{DBLP:conf/emnlp/HayatiOAYTN18} improve code generation by retrieving code similar to the input and copying n-gram actions from the retrieved code. Xu et al.~\cite{DBLP:conf/acl/XuJYVN20} introduces two external knowledge bases from Stack Overflow and API documentation for retrieval and improve model performance. Parvez et al.~\cite{DBLP:conf/emnlp/ParvezACRC21} improve the generation process by introducing similar code snippets alongside the input description into the generator and training the model to selectively incorporate reusable codes.}

\subsection{Program Repair}

Automated Program Repair (APR) methodologies \cite{cacm19} were initially introduced to automatically fix program bugs and reduce the need for intensive manual debugging. In this article we will also examine the possibility of automated repair of automatically generated code.
APR leverages automated techniques to analyze buggy code and generates correct patches to address the identified issues. The research of APR techniques can be mainly divided into three categories: search-based, constraint-based, and learning-based \cite{cacm19}.

\paragraph*{Search-based Program Repair}
Search-based APR methods employ heuristic algorithms to search for the right fix in a predefined patch space~\cite{DBLP:conf/cav/JobstmannGB05}. These methods use heuristics to identify potential bug positions and generate repair candidates. For instance, GenProg~\cite{DBLP:journals/tse/GouesNFW12} uses an extended form of genetic programming to generate program variants that could fix the bugs and retain the required functionalities. RSRepair~\cite{DBLP:conf/icse/QiMLDW14} employs the mutation techniques used in GenProg and uses random search to generate a fix patch. ARJA~\cite{DBLP:journals/tse/YuanB20} formulates automated program repair as a multi-objective search problem and uses NSGA-II~\cite{DBLP:journals/tec/DebAPM02} to look for simpler repairs. One challenge of search-based APR is the costly validation of patches by testing~\cite{DBLP:conf/gecco/ForrestNWG09}. To enhance efficiency, various strategies that try to minimize candidate patches and test cases for validation have been proposed. For example, AE~\cite{DBLP:conf/kbse/WeimerFF13} introduces RepairStrat and TestStrat which leverage equivalent patches to prune semantically-equivalent patches and sample validates patches to cut down costs. relifix~\cite{DBLP:conf/icse/TanR15} targets regression error fixes using previous program versions and contextual repair operators. The work of \cite{DBLP:conf/issta/FryLW12} focuses on the software regression errors and proposes to leverage previous versions of a buggy program.
Search-based repair techniques
may suffer from having to navigate a large search space and to alleviate this issue, fix template guided repair (such as the work of PAR \cite{DBLP:conf/icse/KimNSK13}) has been suggested. Search based repair suffers from
the more serious issue of test-data overfitting where the generated repair can pass the given tests, but not other tests.
The issue of overfitting in program repair, and specifically search-based program repair has been mentioned in \cite{qi2015analysis}
Constraint-based program repair approaches mitigate these concerns, by constructing a
generalization of given tests via symbolic analysis.

\paragraph*{Constraint-based Program Repair}
Constraint-based APR methods utilize constraint specifications to guide the repair by converting the repair problems into a constraint solver problem. For example, SemFix~\cite{DBLP:conf/icse/NguyenQRC13} fixes single-line bugs using symbolic execution by crafting repair constraints. DirectFix~\cite{DBLP:conf/icse/MechtaevYR15} improves patch generation with constraint solving and program synthesis, extending the ability to fix multi-line bugs but suffering from the scalability problem due to maxSMT solving overheads. To overcome this, Angelix~\cite{DBLP:conf/icse/MechtaevYR16} proposes to employ the lightweight value based specifications (angelic forest) for better scalability. Nopol~\cite{DBLP:journals/tse/XuanMDCMDBM17} was proposed to fix if-conditional bugs using SMT. It uses value replacement instead of symbolic execution. Another work called SPR performs enumerative search to find suitable values to be returned by boolean
expressions in different iterations of a loop \cite{SPR15}. The repair tool Prophet \cite{prophet} is an improvement of SPR, where a
machine learning model is employed as the last step to rank patch candidates.


\paragraph*{Learning-based Program Repair}
\textcolor{black}{With the advent of machine learning, numerous methods have been proposed that utilize learning-based models to capture program semantics for repairing bugs. Early deep learning-based APR approaches~\cite{DBLP:conf/aaai/GuptaPKS17,DBLP:conf/wcre/WhiteTMMP19} utilized neural models to learn code semantics for aiding repair tasks, instead of directly generating patches. DeepRepair~\cite{DBLP:conf/wcre/WhiteTMMP19} identifies similarities between buggy code and potential fixes to guide the patch generation. More recent methods~\cite{DBLP:journals/tse/ChenKTPPM21,DBLP:conf/icse/Li0N20,DBLP:conf/issta/LutellierPPLW020} employ neural machine translation (NMT) techniques using encoder-decoder models to understand the semantics of buggy code and translate buggy code into fixed code. For instance, CoCoNuT~\cite{DBLP:conf/issta/LutellierPPLW020} tokenizes code into sequences like text to translate the buggy code into correct code. DLFix~\cite{DBLP:conf/icse/Li0N20} leverages abstract syntax trees with tree-based models to capture code structure information. CURE~\cite{DBLP:conf/icse/JiangL021} integrates pre-trained models in NMT-based APR  and proposes a code-aware search strategy to find compilable patches. Compared with generating patches, Recoder~\cite{DBLP:conf/sigsoft/ZhuSXZY0Z21} proposes to generate the edit to ensure the syntactic correctness of the patched program. The recent work RewardRepair~\cite{DBLP:conf/icse/YeMM22} improves repair performance and the successful compilation rate of patches by training models with program execution information.}

\paragraph*{Security Vulnerability repair} Program repair techniques have shown promise in automatically fixing security vulnerabilities.
This has significant promise and relevance for automatically generated code from LLMs, since security
vulnerabilities in LLM produced code remains a big concern. The work of ExtractFix \cite{extractfix2021} uses address sanitizers to
extract specifications of crash-freedom and then uses symbolic reasoning to produce patches via a repair-augmented
weakest pre-condition computation. This leads to a completely automated vulnerability repair method for memory
errors. The work of SenX \cite{senx19} requires safety properties which are then used to automatically generate vulnerability
patches. Last but not the least, the work of Crashrepair \cite{issta19} suggests a promising workflow where vulnerability
detection via grey-box fuzz testing and vulnerability repair are fused into a single step - prioritizing tests which can
better distinguish among patch candidates. Such workflows may hold promise as automatically generated code from LLMs (potentially replete with security vulnerabilities) become common-place in future.

\subsection{LLM-based Intelligent Programming}

\textcolor{black}{In this section, we will first detail introduce recent representative Large Language Code Models and then introduce some works that utilize LLMs to boost the above programming tasks.}

\subsubsection{Large Language Code Models}

\textcolor{black}{Recently the advent of pre-training techniques techniques has significantly advanced progress in automatic programming. Pre-trained code models are first pre-trained on large-scale unlabeled datasets using self-supervised learning tasks and then fine-tuned or prompted for downstream tasks. Since this process does not require human annotation, it can be applied to large-scale unlabeled datasets, enabling the models to acquire a vast amount of general programming knowledge. Recent studies~\cite{DBLP:journals/corr/abs-2001-08361,DBLP:journals/corr/abs-2206-07682} show that increasing the size of these models significantly boosts their abilities, resulting in substantial enhancements in performance once the models grow beyond a certain parameter threshold. The term ``Large Language Model” (LLM) has been proposed to distinguish these models based on the extent of their parameters. In this section, we will provide a detailed account of well-known Large Language Code Models, ranging in size from Bert-like models to those as large as ChatGPT.}

\textcolor{black}{One pioneer work of pre-trained code model is CodeBERT~\cite{DBLP:conf/emnlp/FengGTDFGS0LJZ20}, which is an encoder-only pre-trained model on six programming languages with two self-supervised tasks, i.e., masked language modeling and replaced token detection, which significantly outperforms previous non-pre-trained models. Another model, CodeT5~\cite{DBLP:conf/emnlp/0034WJH21} is an encoder-decoder pre-trained model following the same architecture as T5. It formulates all the tasks in a sequence-to-sequence paradigm with different task-specific prefixes and achieves promising results on a variety of code intelligence tasks. CodeGPT~\cite{DBLP:journals/corr/abs-2102-04664} is a decoder-only model that pre-trains on programming languages dataset and has the same architecture as GPT-2. PLBART~\cite{DBLP:conf/naacl/AhmadCRC21} uses denoising sequence-to-sequence pretraining for both program understanding and generation purposes. UniXCoder~\cite{DBLP:conf/acl/GuoLDW0022} involves multi-modal contrastive learning and cross-modal generation objective to learn the representation of code fragments. 
More recently, there are also some pre-trained code models that are designed for specific programming tasks such as CodeReviewer~\cite{DBLP:conf/sigsoft/LiLGDJJMGSFS22} and CoditT5~\cite{DBLP:conf/kbse/ZhangP0LG22}.}

\textcolor{black}{Apart from these smaller pre-trained models in academics, many pre-trained code models with much larger sizes have been proposed in the industry in recent years. INCODER~\cite{DBLP:journals/corr/abs-2204-05999} is a model that adopts a causal masking training objective for both code infilling and synthesis and has two versions with 1.3B and 6.7B parameters, respectively. CodeGen~\cite{nijkamp2022codegen} is a large pre-trained model with more than 16B parameters, which achieves promising results for multi-turn program synthesis. Codex~\cite{DBLP:journals/corr/abs-2107-03374} is a large code pre-trained model proposed by OpenAI that supports the service of Copilot. It is adept at understanding and generating code, facilitating the automation of programming tasks, and supporting developers in writing code more efficiently. In addition to Codex, the models recently released by OpenAI, such as ChatGPT~\cite{ChatGPT} and GPT-4~\cite{GPT4}, are also pre-trained on source code data and demonstrate impressive programming abilities.  AlphaCode~\cite{DBLP:journals/corr/abs-2203-07814} is trained for generating code for programming competitions with 715G data and 41B parameters. It can generate novel solutions to unseen programming problems and outperform about half of developers in competitive programming with more than 5,000 participants. StarCoder~\cite{DBLP:journals/corr/abs-2305-06161} is an advanced LLM for assisted programming. Its base version is trained on the Stack dataset with 15.5B parameters and increases the input size into 8000 tokens to enable dealing with longer code. Code Llama~\cite{DBLP:journals/corr/abs-2308-12950} is a family of large-scale code language models developed by Meta and has variations including base, Python-specialized, and instruction-following models, ranging from 7B to 34B parameters. These models are adept at handling sequences up to 100k tokens and are available for both research and commercial use under license. Phi-1~\cite{DBLP:journals/corr/abs-2306-11644}, from Microsoft Research, is a 1.3B parameter decoder-only transformer model, trained on a curated dataset of 7B samples, designed for code-related tasks. WizardCoder~\cite{DBLP:journals/corr/abs-2306-08568} is an open-source LLM based on StarCoder, fine-tuned with instruction-based datasets to enhance code generation capabilities across various complexity levels. DeepSeek Coder~\cite{DeepSeek} is trained on a mixed corpus of code and natural language. It focuses on project-level code completion and infilling and achieves state-of-the-art performance in multiple programming languages on various benchmarks. Magicoder~\cite{DBLP:journals/corr/abs-2312-02120} is a recent work that is trained on synthetic instruction data enhanced with open-source code snippets. Its primary aim is to produce diversified, realistic, and controllable data, addressing the bias typically found in synthetic data generated by LLMs.}

\subsubsection{Utilization of LLMs for Intelligent Programming} 

\textcolor{black}{Recently, apart from training a base LLM, there are also a lot of works that focus on how to utilize these powerful LLMs by tuning or prompting them for automatic programming~\cite{DBLP:journals/corr/abs-2305-06599,DBLP:conf/kbse/GaoWGWZL23,DBLP:journals/corr/abs-2401-01060,DBLP:journals/corr/abs-2304-00385,DBLP:conf/icse/PengGGHL24}.}
In code generation, there is a growing interest in methods that utilize the chain-of-thought prompt to generate better code and solve more complicated programming problems. For example, TIP~\cite{DBLP:journals/corr/abs-2305-06599} utilizes LLMs to formulate a high-level code sketch before working on detailed coding tasks, which improves the precision and reliability of generated code. Dong et al.~\cite{DBLP:journals/corr/abs-2304-07590} proposes a self-collaboration method to advance LLMs in complex coding tasks by employing multiple LLMs as distinct experts and making them interact with each other. Besides, apart from generating codes at function-level, many recent work also explores extending the scope of code generation into into class-level~\cite{DBLP:journals/corr/abs-2308-01861} and repository-level~\cite{DBLP:conf/icml/ShrivastavaLT23}.
As for program repair, there are also a lot of studies utilizing LLMs to repair software bugs. Fan et al.~\cite{fan2023automated} studied the mistakes in auto-generated code and investigated whether existing automated program repair techniques can fix the incorrect code produced by LLMs such as Codex. Xia et al.~\cite{DBLP:conf/icse/XiaWZ23} applied several LLMs for program repair by adopting a infilling-style approach (i.e., predicting what the correct code should look like given
its surrounding prefix and suffix). Huang et al.~\cite{DBLP:conf/kbse/HuangMZLWLZ23} studies the impact of different LLMs and different program repair scenarios. Peng et al.~\cite{DBLP:conf/icse/PengGGHL24} proposes to mine domain-aware fix templates and incorporate them into code prompts to repair Python type error. Apart from the above works that only generate the repair patch in a one-stop way. ChatRepair~\cite{DBLP:journals/corr/abs-2304-00385} leverages the conversational nature of advanced LLMs like ChatGPT and learns from both previous test failure information to provide the model with immediate feedback. With the feedback information from test cases, it could produce more precise and context-sensitive fixes.
Moreover, LLMs are also employed for logging activities such as logging statement automation~\cite{DBLP:conf/icse/Li0CS21,DBLP:conf/icse/ZhuHFZLZ15}. For example, Li et al.~\cite{DBLP:journals/corr/abs-2307-05950} present the first extensive evaluation of LLMs for logging statement generation. Furthermore, Sridhara et al.~\cite{DBLP:journals/corr/abs-2305-16837} explore the proficiency of ChatGPT in summarizing logs, achieving promising results surpassing the existing method. Li et al.~\cite{li2024go} propose to incorporate static context into code prompt and employ a self-refinement manner to further rectify previous errors. Another important field in logging activities is log parsing, which aims at extracting structured templates and parameters from raw log messages to provide insights for developers~\cite{DBLP:conf/icse/ZhuHLHXZL19,DBLP:conf/icws/HeZZL17,DBLP:conf/icse/HuoSLL23}. To facilitate the effectiveness of LLM for log parsing, \cite{DBLP:journals/corr/abs-2307-09950} leverages LLM and in-context learning (ICL) for log template extraction and another work~\cite{DBLP:journals/corr/abs-2310-01796} improves log parsing by ICL and parsing cache. LLMs are also beneficial in generating test cases from natural language descriptions, which enhances cooperation between software developers and testers. These include the automated test case generation of various scenarios such as enhancing the coverage of testing~\cite{DBLP:journals/corr/abs-2305-04764,DBLP:journals/corr/abs-2305-00418} and the detection of possible defects~\cite{DBLP:journals/corr/abs-2305-04764}. Ryan et al.~\cite{DBLP:journals/corr/abs-2402-00097} proposes to provide LLMs with path constrains and code context to improve the coverage of generated test cases. ChatUniTest~\cite{DBLP:journals/corr/abs-2305-04764} extracts essential information and creates an adaptive focal context for LLMs to generate test cases.

\section{Program Repair and Auto-coding}

Program synthesis converts a formal or semi-formal specification into expressions or code snippets. The area has been studied as early as \cite{pnueli} and a recent survey appears in \cite{alur}. The specifications driving program synthesis may often be given as a collection of {\em (input, output)} examples - providing the oracle for a given input. Program repair \cite{cacm19} involves a correction or rectification of a code-base so that it can meet certain correctness criteria. The correctness criteria can be given in terms of system level test cases that the overall software system needs to pass.
Both program synthesis and repair suffer from the {\em overfitting} problem due to the incompleteness of the specifications driving these processes. If the specification is given as a test-suite the overfitting can appear in the form of the generated code overfitting the test-data.  As a simple example let us suppose we have {\em (input, output)}  specifications given in terms of collections of input-output pairs as follows.
\begin{verbatim}
(input = 2, output = 4)
(input = 3, outout = 9)
\end{verbatim}
and we have a buggy program 
\begin{verbatim}
output = input + input;
\end{verbatim}
An inadequate program repair system may fix the above program to 
\begin{verbatim}
if (input == 2)  output = 4;
else if (input == 3) outout = 9;
\end{verbatim}
while our desire will be to produce the following (minimal) fix via program repair 
\begin{verbatim}
output = input * input;
\end{verbatim}

This simple example also makes it apparent the core issue of "generalization" underlying program synthesis approaches - particularly those that are driven by input-output examples. It is always possible to synthesize code that works exactly for the given input-output examples by producing code with the following schematic
\begin{verbatim}
    if (input == input1) return output1
    else if (input == input2) return output2
    else ...
\end{verbatim}
Imposing certain quality indicators such as code size may induce the program synthesizer to produce more compact code which {\em generalizes} the given input-output examples. 
While there exist a large number of synthesis approaches, many of them typically perform an enumerative search over the search space of expressions. The enumerative search may be guided by a choice of operators appearing in the expression (component-based synthesis \cite{jha2010oracle}) or certain restrictions over the syntax of expressions typically captured via a grammar (syntax-guided synthesis \cite{sygus}). Irrespective of the technical machinery used to conduct the synthesis - the issue of overfitting of the synthesized code remains. The concern is that the synthesized code may return the expected output for the given input-output examples but not for other inputs. For the program synthesis problem, this problem sometimes remains implicit - since the expected output for inputs other than those appearing in the given (input, output) examples may not even be documented fully. In the problem of program repair, where a buggy program is given - the problem of overfitting is more explicit. Here the fixed program may pass the given tests in a test-suite which is used to guide the repair; at the same time, the fixed program may not pass tests outside the given test-suite.

We now discuss a treatment of program repair as a field with some technical glimpses on the underlying challenges such as over-fitting. {\em The treatment is from the open-source unpublished article by the third author \cite{repair-arxiv}.}

\subsection{Program repair}

The issue of overfitting has been well studied and articulated in the area of program repair \cite{qi2015analysis}. While raising awareness about the issue, these works have articulated concerns which go beyond the incompleteness of tests. It is generally known that any test-suite as collection of (input, expected output)  pairs is an incomplete specification of intended program behavior. Therefore repairs generated by using a test-suite $T$ as guidance may not pass tests outside $T$. However the concerns about generating overfitting repairs go beyond the incompleteness of $T$. For example if the oracle of certain tests say that an exception should not raised, a repair may simply delete the code which raises these exceptions and meet the requirement.  

For this reason, it is important for automated program repair techniques to 
\begin{itemize}
\item perform an adequate generalization of the given test-suite $T$, so that the repairs do not only work on tests in $T$ 
\item satisfy certain code quality indicators apart from passing the given tests, to avoid obviously unacceptable repairs such as deleting the code checking the oracle. 
\item to ensure quality patches certain repair techniques emphasize the succinctness of the patches - meaning smaller disruption to the code-base is somehow "better".
\end{itemize}

We now describe in details one concrete approach for program repair, which seeks to achieve these goals by symbolic analysis of the given tests in $T$. Here symbolic analysis of the test executions for tests in $T$, amounts to computing a generalization which we want to work for tests outside $T$ as well. 
By symbolically analyzing the tests in $T$, the repair method extracts specifications about the patch code in the form of {\em repair constraints}. These repair constraints can be used as guidance in generating patches via search or program synthesis. We emphasize here for the reader that this is only one approach for program repair, and there exist several other approaches based on search and learning \cite{cacm19}. One motivation for presenting this constraint based program repair approach is to illustrate ideas about how automatically generated code in program repair techniques can avoid the test over-fitting problem. Conceptually speaking, we could always define a domain of program edits, and then conduct a random search in this domain to find edits which pass given tests. However, the output of such a search would be greatly dependent on the search heuristics and it would be hard to give any assurance about the quality of the patches. Thus, instead of searching at random in the space of patches - we show how the repair technique can be "guided" to produce higher quality patches. 

In the approach that we elaborate in prior work~\cite{semfix}, the repair technique is "guided" by a repair constraint which generated by symbolically executing the tests in the given test-suite $T$ in a novel fashion. So, the main conceptual step is in using constraints to reduce the search space of possible patches, as opposed to searching in the domain of patches. We do not discuss the computation of the constraint in details, but rather conceptualize at a high level how the presence of such a constraint can help generate high quality repairs and avoid patch overfitting. 

\begin{figure}
\begin{tabular}{cc}
\begin{tabular}{cl}
1   & {\tt int tri\_detect(int a, int b, int c)\{} \\
2   & \ \ \ {\tt if (a <= 0 || b <= 0 || c <= 0)} \\
3   &  \ \ \ \ \ \ {\tt return INVALID;}\\
4   & \ \ \ {\tt else if (a == b \&\& b == c)}\\
5   & \ \ \ \ \ \  {\tt return EQUILATERAL;}\\
6   & \ \ \ {\tt else if (a == b || b == c)}\\ 
7   & \ \ \ \ \ \  {\tt return ISOSCELES;}\\
8   & \ \ \ {\tt else return SCALENE;}\\
9   & {\tt \}}
\end{tabular} & \hspace*{0.2in}
\begin{tabular}{|c|c|c|c| c|}
    \hline
    a & b & c & Output & Outcome\\
    \hline\hline
    -1 & 1 & 1 & INVALID & Pass\\ \hline
    2 & 2 & 2 & EQUILATERAL & Pass \\\hline
    2 & 2 & 3 & ISOSCELES & Pass \\ \hline
    2 & 3 & 2 & SCALENE & Fail \\ \hline
    3 & 2 & 2 & ISOSCELES & Pass \\ \hline
    2 & 3 & 4 & SCALENE & Pass \\ \hline
\end{tabular}
\end{tabular}
    \caption{Triangle program from \cite{cacm19} and the test data accompanying the program}
    \label{fig:my_label}
\end{figure}

Let us consider a program that takes in three sides of a triangle and determines the kind of triangle constructed out of these three sides. The program may look like the program in Figure \ref{fig:my_label}. 
This program has several bugs. For three sides which violate the triangle inequality - it should return INVALID, but it is not doing so. Similarly, the definition of the isosceles triangle is supposed to check if any two of the three sides are equal. Now, as shown in 
the test suite from Figure~\ref{fig:my_label},let us show a realistic test-suite consisting one test for invalid triangle, one for equilateral triangle, three tests for isosceles triangle (depending on which two sides are equal), and one test for a scalene triangle. A reasonably constructed test-suite based on the requirements will indeed be of this nature.
Let us assume now that by a control flow analysis of the passing and failing tests, line 6 is inferred as the fix location.  The fix localization process is the same as the localization of search-based APR techniques.
The exact process of fix localization is not shown here. It may involve finding out locations which appear with significantly greater frequency in failing tests, than in passing tests. Once the fix location is identified, the expression in that location is substituted as an unknown or a symbolic variable X. 

\begin{verbatim}
...
6       else if (X)
7           return ISOSCELES;
...
\end{verbatim}

Now, it is required to find out properties about X which would make the program pass the test cases that are given. 
\begin{itemize}
    \item the first two tests do not even reach line 6.
    \item among the remaining four tests that reach line 6, X should be true in the third, fourth, and fifth tests. Moreover, X should be false in the sixth test.
    
\end{itemize}
Getting the above-mentioned requirements, though put intuitively here, is not straightforward. It involves an analysis of the test executions for the given tests. Essentially it amounts to finding the desired value of X (in this case a boolean as it represents a boolean expression) so that it can make the test pass. This is captured by the {\em repair constraint}.

How to formally capture these requirements or constraints on X, which essentially is a placeholder for the code inserted in line 6? A formal way of understanding this repair constraint is that the unknown X is essentially an unknown function on the variables which are live in line 6. Thus essentially
\begin{equation}
\begin{split}
X = f(a,b,c)
\end{split}
\end{equation}
where $f$ is an unknown function that is to be synthesized. The information about the function $f$ is given by the following repair constraint.
\begin{equation}
\begin{split}
f(2,2,3) \wedge f(3,2,2) \wedge f(2,3,2) \wedge \neg f(2,3,4)  
\end{split}
\end{equation}

This repair constraint can be fed to a program synthesis engine. The synthesis engine can be fed with the ingredients that can appear in the expression: the variables, the constants, and the operators. In this case, the variables are {\tt a}, {\tt b}, {\tt c}, the constants are the integer constants and the operators are the relational operators and logical operators. With these ingredients and the provided repair constraint, a component-based synthesis engine \citep{oracle-guided10} will yield the correct fix 
\begin{equation}
\begin{split}
    f(a,b,c) = (a == b || b ==c || a == c)
\end{split}
\end{equation}

Let us now present the formal treatment of repair constraint computation. Statistical fault localization \citep{SBFL-Survey} or other offline analysis techniques are applied to identify potential fix locations. Such offline analysis may involve program dependency analysis, or simply control flow analysis of the passing / failing tests. Let us examine how the control flow analysis of passing / failing tests will proceed under the auspices of statistical fault localization. In such an approach, each statement $s$ in the program is given a suspiciousness score based on the occurrences of $s$ in the passing / failing tests. Constraint-based APR techniques also rely on fault localization to determine the line to be fixed.
Once a fix line is decided, a repair constraint is then constructed. This is a constraint on the expression to be put in the corresponding line as a fix. For the purposes of explanation, let us assume that the fix is either a boolean expression or an arithmetic expression which is the right hand side of an assignment. How to construct the repair constraint? For a boolean expression, the expression can be simply replaced with a new symbolic variable {\tt X} as follows.
\begin{equation}
\begin{split}
    {\tt if(e)} \rightarrow {\tt if(X)}
\end{split}
\end{equation}
For an arithmetic expression, a new symbolic variable {\tt X} is introduced as follows.
\begin{equation}
\begin{split}
    {\tt y = e} \rightarrow {\tt y = X}
\end{split}
\end{equation}
Note here that {\tt y} is a program variable and {\tt e} is an expression made out of program variables, while {\tt X} is a symbolic ghost variable which is introduced by us, for the purposes of automated program repair. Note that the symbolic variable {\tt X} is introduced at the deemed fix location, and for now let us assume we are generating a one line fix.

\begin{figure}[!t]
    \centering
    \includegraphics[width=0.55\textwidth]{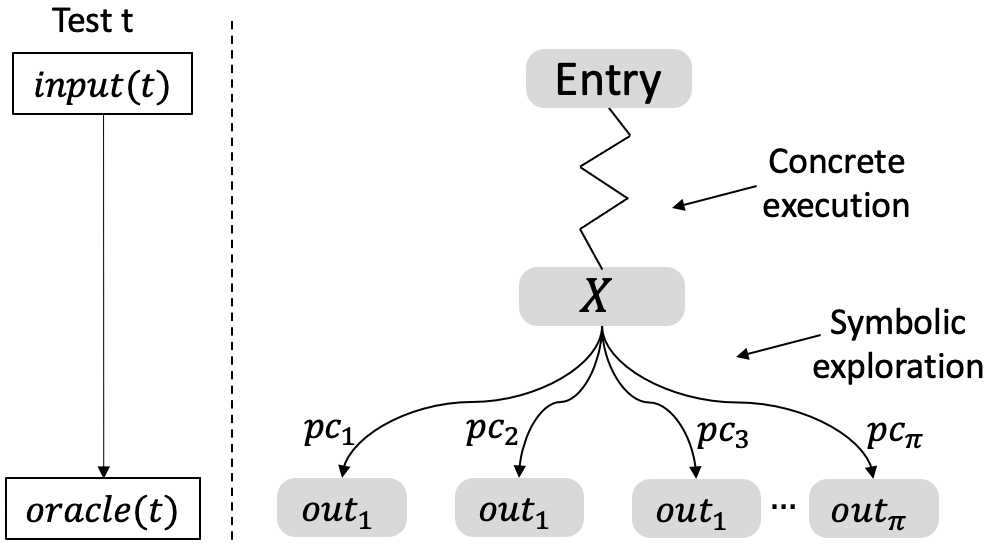}
    \caption{Inferring Specifications for Program Repair (ack. unpublished article \cite{repair-arxiv})}
    \label{fig:repair_constraint}
\end{figure}

Given such a ghost symbolic variable {\tt X}, the repair constraint is defined in terms of {\tt X} as follows. For a given test $t$, the path up to the fix location $L$ is concrete. From the fix location $L$, there are several possible paths, depending on the value of {\tt X}. Therefore, the path condition of a path $\pi$ from $L$ and the symbolic output along the path in terms of $X$ can be defined. Let these be $pc_\pi$ and $out_\pi$ respectively, as illustrated at Figure~\ref{fig:repair_constraint}. Then a constraint for path $\pi$ can be represented as 
\begin{equation}
\begin{split}
    pc_\pi \wedge out_\pi = oracle(t)
\end{split}
\end{equation}
where $oracle(t)$ is the expected output for test case $t$. Considering the various paths from $L$ for the execution of test $t$, repair constraint for test $t$ to pass is 
\begin{equation}
\begin{split}
    C_t \equiv \bigvee_{\pi} pc_\pi \wedge out_\pi = oracle(t)
\end{split}
\end{equation}
The overall repair constraint is the conjunction of the repair constraint collected from all the given tests, since the repaired program is expected to pass all the given tests.
In other words, the repair constraint C is given as follows.
\begin{equation}
\begin{split}
    C \equiv \bigwedge_{t} C_t
\end{split}
\end{equation}

\smallskip

\subsection{Language Model based Code Generation}
Designing AI-based systems to automatically solve programming tasks has gained considerable attention in recent years.
The most notable of these comes in the form of transformer-based large-scale language models, which used to transform natural language text.
Large language models, such as Codex~\citep{codex} and AlphaCode~\citep{alphacode}, have also successfully generated code for many programming tasks in Python, Java, C, etc.

\paragraph{Program Repair for fixing Code Generated by Language Model}
Codex and AlphaCode have shown capability in generating correct solutions for many programming tasks.
However, the success rate of existing language models remains low, especially for complex programming tasks.
One of the reasons is that language models lack awareness of program semantics (e.g., type information, run-time program states, etc.), resulting in incorrect programs.
A large part of bugs made by Codex are syntax errors or misaligned algorithms, i.e., uncompiled programs or programs with incorrect algorithms.
Meanwhile, some bugs require small changes, e.g., changing operators, modifying expressions, or changing statements.
For instance, Figure~\ref{figure:codex} shows an example program produced by Codex for an programming task in LeetCode~\footnote{https://leetcode.com}.
The comments in Figure~\ref{figure:codex} are the program descriptions, which are provided to Codex as prompt, and the code is automatically generated by Codex.
Unfortunately, the produced program has a bug, causing the program to fail on some test cases.
The correct fix is to change statement $i {-}{=} 2$ at line 9 to $i {-}{=} 1$.
Compared to language models, typical repair
tools generate patches by reasoning about the program semantics against the given specification.
Hence, the repair technique has the potential to increase the success rate of language models.
In the above example, several existing repair tools can automatically fix the bug and make it pass all the test cases.

\begin{figure}[!t]
\centering
\begin{verbatim}
1  //A fancy string is a string where no three consecutive characters are equal. 
2  Given a string s, delete the minimum possible number of characters from s to make it fancy. 
3  Return the final string after the deletion. It can be shown that the answer will always be unique.
4  public String makeFancyString(String s) {
5  StringBuilder sb = new StringBuilder(s);
6   for (int i = 2; i < sb.length(); i++) {
7     if (sb.charAt(i) == sb.charAt(i-1) && sb.charAt(i) == sb.charAt(i-2)) {
8         sb.deleteCharAt(i);
9  -      i -= 2;
10 +      i -= 1; 
11     }
12    }
13   return sb.toString();
14  }
\end{verbatim}
\caption{The program for a LeetCode programming task generated by Codex.}
\label{figure:codex}
\end{figure}

\paragraph{Language Model for Program Repair}
Language models could also be used for fixing software bugs. In March 2022, a new version of Codex edit mode was released. Instead of just translating program descriptions to programs~\footnote{https://openai.com/blog/gpt-3-edit-insert}, the Codex edit model can change existing code in a complete program.
This new feature makes it practical to use Codex for program repair.
Codex edit mode requires users to provide instructions to guide the code change, such as ``fix the bug at line 2'', or ``fix the index-out-of-bound exception''.
To fix a bug, users need to provide precise and clear instructions.
The repair based on large language models could even produce better performance in fixing software bugs than learning based repair techniques.
Compared to existing learning-based repair, e.g., SequenceR and Recoder, Codex is trained on a much larger dataset than Recoder, which helps Codex to learn more fix patterns (see \cite{fan2023automated} for comparison results).
In fact, large language models learn code edit patterns from huge existing programming artifacts (including code, commits, comments and etc.).  Nevertheless the task of repairing automatically generated code remains a challenge. A recent work \cite{fan2023automated} has shown that automatically generated code for even simple Leetcode problems contain large number of occurrences of ``{\em misaligned algorithms}".  Such misaligned algorithms constitute cases where the programming of the task (or one of its sub-tasks) is incorrect even at an algorithmic level - so that the code needs to be completely re-written.


\paragraph*{Using the latest LLMs}
Prompting some of the largest commercial models, such as Gemini 1.5, GPT3.5, and GPT4, shows that the example generated by Codex can be successfully repaired. Providing no specific information about the bug ("fix the issue please") or only the buggy line ("there is an error on line 9") is sufficient to get a patch. Furthermore, these models are commonly used for chatting and generate a human-readable explanation of the fault. Figure \ref{figure:gpt4} is an example of such an interaction with OpenAI's GPT4 model. The flexibility and power of such models come at a (small) cost. With an imprecise prompt, the model may respond in an unexpected format, which cannot be used to repair the program automatically. A solution is to describe how the model should return the fix. An example modification is appending the string ``Return the patched code in triple backticks and nothing else.''
With the advancement of the models' size and architecture, we postulate that syntax errors for common languages will occur less in the output of large commercial models in the coming years. On the other hand, as natural language is inherently ambiguous, the issue of misaligned algorithms will remain problematic. Misaligned algorithms occur when while trying to write code for a task, a language model generates an incorrect solution even at algorithmic level for some of the sub-tasks \cite{fan2023automated}.
Validation techniques must be defined to alleviate this, and guarantees of the model's response must be provided to ensure safe code integration.
\begin{figure}
\begin{verbatim}
The given code snippet aims to remove characters from the string `s` 
to ensure no three consecutive characters are the same, thereby making it a "fancy string." 
However, there's an issue with the logic that might not always yield the correct result 
because of how the index `i` is managed after deletion. When you delete a character at index `i`, 
you do not need to decrement `i` by 2. Instead, you should decrement `i` by 1 to ensure the loop
checks the next set of characters properly after a deletion.

Let's correct the code:

...

Here's the corrected version:

```java
public String makeFancyString(String s) {
    StringBuilder sb = new StringBuilder(s);
    for (int i = 2; i < sb.length(); i++) {
        if (sb.charAt(i) == sb.charAt(i-1) && sb.charAt(i) == sb.charAt(i-2)) {
            sb.deleteCharAt(i);
            i--; // Corrected to decrement i by 1 instead of 2
        }
    }
    return sb.toString();
}
```
...
\end{verbatim}
\caption{Example repair generated by GPT-4 (abbreviated)}
\label{figure:gpt4}
\end{figure}

\paragraph*{Commentary on semantic approaches}
We postulate that language model based code generation, as well as language model based repair approaches could play a role in future. At the same time, the relationship of the language model based repair approach with respect to program synthesis is not well-understood today. Since semantic repair approaches or constraint based repair approaches, rely on symbolic reasoning, there exist opportunities in combining semantic repair approaches with language model based repair in the future. Here we need to be careful about what kind of back-end we use for the constraint-based repair approach. {\em We note that the program synthesis back-end can be replaced by a generative AI model}, while the constraint-based repair can provide a systematic selection mechanism for selecting among patch candidates. These workflows can be examined in the future.

\section{Code LLMs, Software Quality and Trustworthiness}
\label{sec:quality}

The use of LLMs in automatic programming shows immense potential. However, this progress brings with it concerns about the trustworthiness of the code these models produce. For instance, Jesse et al.~\cite{jesse2023large} found code LLMs tend to introduce simple bugs in the codebase. Estimating correctness of the code generated by LLMs becomes problematic, especially in the absence of clear specifications. 
In fact, there have been instances where code generated by LLMs was found to contain security vulnerabilities~\cite{pearce2022asleep, perry2023users}. Identifying security flaws within the generated code is challenging; developers might not review every piece of code in detail, leading to overlooked errors. 
There is also a risk of LLMs being exploited by bad actors who may tamper with the training data or manipulate the prompts used during the query phase~\cite{greshake2023more}.
The opaque nature of LLMs adds another layer of complexity to the task of analyzing and debugging automatically generated code. The intricate algorithms that drive the code generation process are not fully transparent to developers, making it hard for them to grasp how the code comes to exist. This issue becomes even more pronounced in programming environments where code is continuously edited. The need for an Integrated Development Environment (IDE) that can not only generate code efficiently but also provide clear, non-intrusive explanations is critical.

\subsection{Quality}
To systematically access the software quality of automatically generated code by LLMs, we refer to  ISO/IEC 25010 guidelines. Specifically, ISO/IEC 25010 includes eight quality characteristics: (1) functional suitability (i.e., functional completeness, functional correctness, and functional appropriateness), (2) performance efficiency, (3) compatibility, (4) usability, (5) reliability, (6) security, (7) maintainability, and (8) portability. Overall, we notice that most of the recent research focuses on studying functional suitability~\cite{fan2023automated}, usability~\cite{vaithilingam2022expectation}, reliability~\cite{poesia2022synchromesh,zhong2023study}, security~\cite{perry2023users}, and maintainability~\cite{liu2023refining}. From these existing studies, we derive a few observations. Firstly, despite the recent advancement in LLMs like ChatGPT, LLMs still generally produce low-quality code based on recent evaluations that cover different quality characteristics. Secondly, \emph{there exists a recent trend of studies to cover more quality characteristics by proposing new benchmarks}. For example, LMDefects~\cite{fan2023automated} contains LLM-generated programs that are functionally incorrect, whereas NoFunEval benchmark~\cite{singhal2024nofuneval} has been recently proposed to evaluate non-functional requirements including performance efficiency and security and the study on the benchmark has noted the low performances of code generation models in non-functional requirements. Thirdly, existing studies rely on traditional metrics for human-written code to access the quality of automatically generated code and these \emph{traditional metrics are still generally applicable for automatically generated code}. For example, prior study relies on static analysis tools for measuring maintainability~\cite{liu2023refining} and found that ChatGPT-generated code suffer from maintainability issues.

An experimental evaluation of Large Language Models for code appears in \cite{Chen21,Xu22} and we refer the reader to these articles. The evaluation includes the {\em Pass@1} rate for the code generated from these models, which is the percentage where a randomly generated code sample passes all given unit tests.

Despite many recent studies on LLM-generated code, we notice that \emph{quality characteristics such as compatibility, and portability are still under-explored}. Notably, existing studies mostly focus on compatibility issues related to test scripts generation~\cite{yu2023llm} and library-related issues~\cite{liao2023context}. For example, a recent approach~\cite{liao2023context} proposed to add awareness of third-party library information to improve the accuracy and the reusability of the generated code. As adding awareness of a quality characteristic have shown promising results in improving the quality of generated code~\cite{liao2023context}, one viable solution would be to fuse all the eight quality characteristics into LLM to improve the overall quality of the generated code. However, as some of the characteristics may have conflicting requirements, one viable solution is to guide LLM to prioritize certain quality characteristics for different tasks or applications. For example, as secure code may be less efficient due to the additional security check, we need to encode the priority for  security over performance efficiency when using LLM to generate code for certain safety-critical systems. Another viable solution is to define a set of anti-patterns~\cite{anti-pattern} for various quality characteristics and encode these ``bad patches'' into LLMs as rules to improve the quality of the generated patches.

We now discuss the more specific issue of {\em trustworthiness} of LLM generated code, and what it would take to trust the integration of LLM generated code as methods into our software project.

\subsection{Trustworthiness in integrating LLM generated Code}
\label{sec:trust}

\begin{table}
    \centering
    \begin{tabular}{lll}
    \toprule
    Dimension & Explanation \\
    \midrule
      \multirow{3}{*}{Code-specific} & Security       & The code generated by LLMs should not have any security vulnerabilities. \\
       & Reliability    & LLM generated code should be free of bugs.\\
       & Privacy        & Code LLMs will not leak unauthorized information. \\ \midrule
      \multirow{5}{*}{Model-specific}  & Explainability & The model should be able to explain its rational of producing certain code or decision.  \\
       & Robustness     & Code LLMs should maintain their performances under diverse noisy inputs. \\
       & Consistency    & The models' outputs should be consistent and reproducible. \\
       & Fairness       & The model should not produce any code or decision exhibiting unethical or unfair behavior.\\
       & Ethics         & The model should not produce any code that intentionally causes harm to humanity. \\
    \bottomrule
    \end{tabular}
    \caption{Attributes of Trustworthiness for Code LLMs}
    \label{tab:dimTrust}
\end{table}

In the near future, ensuring the seamless integration of LLM-generated code into real-world code bases with greater reliability will be crucial. It is imperative to delve into the concept of trustworthiness specifically concerning code generated by LLMs and to develop systematic methods for evaluating this trustworthiness. This exploration will not only inform the future development of models but also shape the entire Software Engineering ecosystem surrounding them. By understanding and addressing these aspects, we can pave the way for more robust and dependable utilization of LLMs in coding applications.


To this end, we reviewed existing literature on the trustworthiness of software~\cite{becker2006trustworthy, schneider1999trust} and trustworthiness of generic LLMs~\cite{sun2024trustllm}. None of them individually is sufficient for our purpose.  Drawing upon this research, we identified eight primary attributes essential for evaluating the trustworthiness of code LLMs, as outlined in Table~\ref{tab:dimTrust}. These attributes can be broadly categorized into two main groups: (i) those pertaining to the properties of the generated code  and (ii) those broadly relevant to LLMs but they can be adopted for Code LLMs. By delineating these attributes, in the future, it will be essential to establish a comprehensive framework for assessing the trustworthiness of code generated by LLMs, thereby facilitating informed decisions regarding their utilization in real-world applications. In the following paragraph, we will elaborate on this, especially for code related attributes.


\paragraph{Security} Given that LLMs are trained on a vast amount of open-source code corpus, it is likely that the pre-trained code corpus contains unverified code that contain security vulnerabilities. The LLMs learn from such vulnerable and exploitable examples. This raises concerns about the security of the code it generates. To check this issue, Pearce et al.~\cite{pearce2022asleep} methodically examine the prevalence and circumstances under which GitHub Copilot might generate insecure code. Their analysis involves prompting Copilot to generate code in situations relevant to high-risk cybersecurity vulnerabilities, such as those identified in MITRE's "Top 25" Common Weakness Enumeration (CWE) list. They came up with 89 unique scenarios for Copilot to tackle, resulting in the creation of 1,689 programs. Among these, approximately 40\% were found to be vulnerable to exploitation. Similar observations were found in other independent studies~\cite{perry2023users, asare2023github}.

\paragraph{Reliability.}  Code LLMs tend to produce subtle trivial bugs as well. In fact,  Jesse et al.~\cite{jesse2023large} reported that Codex and other LLMs produce verbatim single statement bugs up to twice as often as known, for verbatim correct code.

\paragraph{Privacy.} A substantial amount of code data, which is required to train these models, becomes a hindrance, as companies are understandably hesitant to share such sensitive data. This reluctance stems from the fear of potential leaks of proprietary information, including sensitive information like names, emails, passwords, etc~\cite{niu2023codexleaks}. Even when considering third-party foundation LLMs, companies remain cautious about exposing their code to external entities. One of the central challenges in this context revolves around harnessing the capabilities of LLMs while ensuring the protection of proprietary information. Striking a balance between leveraging the power of these models for software engineering tasks and safeguarding sensitive data poses a significant hurdle that the research community must navigate. As the use of LLMs becomes more prevalent, addressing these security and privacy concerns will be crucial to realizing their full potential in the field of software engineering.

\paragraph{Potential Remedy.} When incorporating code generated by LLMs into projects, it is crucial to ensure that the generated code is free from obvious vulnerabilities, errors, or leaks of sensitive information. Integrating the checks as part of the automated development process will be even more important for maintaining the security and integrity of the software. To achieve this, we outline few strategies: 

\begin{itemize}
    \item Firstly, we should prioritize using high-quality training data for the LLMs. This entails training the models on datasets that are thoroughly vetted and free from known vulnerabilities or bugs. By starting with clean and reliable data, the likelihood of the LLM generating flawed code can be significantly reduced.
   \item Additionally, developing and employing light-weight static analysis tools can be instrumental in evaluating the quality of the code generated by LLMs. These tools can automatically analyze the code for potential vulnerabilities, syntax errors, or other issues without the need to execute the code. By running static analysis on the LLM-generated code, developers can identify and address any issues before integrating it into their projects.
   Note that, such static analysis tools should be light-weight and fast as they need to be integrated with IDE and should not hinder developers' productivity significantly. The static analysis should be able to analyze even partial programs, as the code generated in the IDEs may not be complete. Further, last mile improvement of code generated from LLMs, can be enabled by automated program repair. 
   \item To boost LLM intelligence and reliability, integrating step-by-step logical reasoning and self-debugging abilities is key. This equips models to better grasp code context, resulting in more accurate outputs. Self-debugging empowers LLMs to automatically identify and fix errors, potentially reducing vulnerabilities. These enhancements enhance overall model reliability, leading to higher-quality, secure code.
   \item Last but not the least, there exist enticing possibilities of generating verified code with the help of Large Language Models. This can take many forms, including (a) generating code from LLMs and systematically improving it to produce verified code, or (b) generating code in a verified programming language. We note that some efforts along these lines have already been started, such as \cite{Misu24} reporting the LLM-assisted synthesis of verified Dafny methods.
\end{itemize}

\section{Programmer-LLM interaction}



Given the increasing capabilities of AI, particularly Large Language Models (LLMs), in automatic programming, there is a surge in the development and integration of tools based on code-fluent and LLMs to serve as programming assistants.
This section provides an overview of how humans can engage with AI models and Large Language Models (LLMs) for automatic programming. Specifically, we highlight two main common interaction patterns: {\em Autocompletion} and {\em Prompting}.
We then discuss the challenges that programmers face when leveraging LLMs.

\subsection{Interaction Patterns}  
\paragraph*{Autocompletion} 
Autocompletion for code refers to the seamless integration of AI models into an Integrated Development Environment (IDE) without requiring explicit user invocation.
Generally speaking, this tool continuously queries the AI model for code suggestions and promptly displays them to the user. Users engage with the AI model by selecting and validating the generated suggestions. A notable example is GitHub Copilot~\cite{Bird2023}, which acts as an AI pair programmer. Copilot takes preceding code comments or source code (e.g., a function header or partial implementation) as input. It then offers suggestions to complete the remaining implementation whenever the user pauses.
One outstanding advantage of AI-based autocompletion is its ability to complete multiple lines of code in one suggestion. This ability significantly improves usability compared to traditional completion tools, which typically suggest one subsequent token at a time.
With this advantage, programmers can also utilize the tool as a substitute for internet searches \cite{Vaithilingam2022}. 
This would reduce cognitive load as programmers can focus on tasks within the IDE.


\paragraph*{Prompting:} Instead of relying on AI models to infer tasks from code comments or preceding source code, programmers explicitly provide specialized input called prompts that provide instructions on how the LLMs should generate code. 
With the interaction pattern of explicit invocation by programmers, there are various ways that programmers can interact with the AI models.
For example, GenLine \cite{Jiang2022} provides a command-like interaction style where programmers specify a command (e.g., \texttt{“[[html: make an OK button]]”}) within the code to invoke the AI models.
Alternatively,  AI models can serve as virtual coding assistants within the IDE, allowing programmers to provide instructions through a dedicated user interface like a textbox \cite{Kazemitabaar2023}.
With this kind of interaction, programmers can provide structured instructions (e.g., chain-of-thoughts) as a prompt. To enable more engagement with the AI models, programmers can engage with the AI models via a conversational interaction where it takes the previous invocation as the additional context of the input prompts \cite{Ross2023}. Two common intentions of developers to use these tools are 1) acceleration and 2) exploration \cite{Barke2023}. In the acceleration mode, programmers intend to use the models when they have specific programming tasks in mind in mind and leverage the AI model to promptly complete them instead of typing.  The tasks are typically small and logical subtasks that demand less analysis and more straightforward coding, often perceived as just tedious work. Thus, to harness the acceleration potential of the AI models in coding, it is essential to first analyze and decompose the complex task into smaller logical subtasks. The exploration mode emerges when programmers encounter a new problem and are uncertain about how to decompose the task. As the tool can take code comments which is a natural text describing programming intent to generate suggestions, programmers to craft various code comments as inputs and then explore multiple implementation suggestions. Even if the suggestions are not entirely correct, they may still provide a code skeleton or starting point \cite{Vaithilingam2022}. Alternatively, in the exploration mode, programmers use AI-based autocompletion instead of searching for solutions on the internet or StackOverflow \cite{Barke2023}.



\subsection{Usability Challenges}
While LLMs have shown promising results in coding assistance, offering programmers prompt completion of implementations or opening new avenues for exploring alternative programming solutions, new challenges emerge when programmers interact with them.

The first challenge lies in crafting the input. The models take a natural language text (e.g., code comments, prompt) which describes programming intentions as inputs.
Several studies found that the AI models are sensitive to these inputs.
A slight deviation can result in significantly different code generation \cite{Vaithilingam2022,Denny2022}.
Hence, programmers may need to spend time exploring, crafting, and revising inputs to generate correct solutions \cite{Jiang2022}.
Occasionally, programmers had to write extensive and detailed descriptions for the inputs to make the models generate correct solutions. However, this process may consume more time compared to writing the code directly \cite{Kazemitabaar2023}.
With the current interactions with AI models, the ability to create effective prompts may potentially become an important skill in programming.

The second challenge centers around understanding and validating the generated code.
Since the code is generated by the AI models, programmers' main focus has shifted from programming to assessing the suggestions. 
Unlike traditional code completion tools, which usually suggest one subsequent token at a time, LLMs can generate a lengthy sequence of tokens to complete the entire implementation. Consequently, understanding the generated code and validating whether it aligns with programming intentions could demand considerable time and cognitive load \cite{Tang2023, Jiang2022}. 
\citet{Barke2023} found that programmers tend to look for the presence of certain keywords
or control structures to quickly validate suggestions.
They may also execute the code or run a static analyzer to help them validate the suggestions. Herein also lies our hypothesis that with the arrival of LLM-based coding, the nature of {\em program comprehension} activity is likely to shift from manual code comprehension to an iterative dialogue with LLMs. The first step of such an iterative dialogue is of course a validation or disambiguation of artifacts produced by LLMs. 

The third challenge involves debugging and fixing the generated code. 
Even though programmers can understand the generated code, it might still require fixing or improvement. 
However, the AI model may generate complex code that is difficult to debug \cite{Barke2023}. 
Programmers also need to consider the time required in debugging and fixing the generated code; otherwise, they might get stuck in a time-consuming debugging process \cite{Vaithilingam2022}. Additionally, constant context switching between programming and debugging modes can impose significant mental demands on programmers.


\subsection{LLMs for Maintenance \& Evolution}


LLM and AI-based code models have demonstrated significant advancements in expediting coding within automatic programming. It is crucial, however, that automatic programming not only focuses on accelerating the coding process but also contributes to software maintenance and promotes future evolution. 
Considerable effort has been devoted to developing approaches for LLMs to achieve this goal in various ways. 
In this section, we will discuss AI-based approaches where source code is taken as input to generate non-source code
artifacts that facilitate maintenance and evolution.
Specifically, we focus on three main tasks that are closely related to the programming task, i.e., code summarization, code change summarization, and code review.

\paragraph*{Code Summarization}
Code summarization refers to the summarizing of the behaviour or purpose of the provided code snippets~\cite{Ahmad2020,Wu2020}. This is particularly useful for developers when they need to understand the source code, especially the code they have not written themselves.
Recent LLMs like GPT, Codex, CodeT5, CodeBERT, UniXCoder have been investigated for code summarization purposes as these models were trained with multimodal data~\cite{Ahmed2023, Arakelyan2023,DBLP:conf/emnlp/0034WJH21,Gao2023,Gu2022}. 
Thus, the models can generate natural language descriptions from source code.
Some studies also found that the performance of LLMs can be improved when the models learn few-shot exemplars (a.k.a in-context learning)~\cite{Ahmed2023,Gao2023}.

\paragraph*{Code Change Summarization}
Code change summarization refers to the process of summarizing a collection of code changes (e.g., commits or pull requests) made to the codebase. 
It involves describing an overview and purpose of the changes.
This task is crucial for developers as it helps developers and other stakeholders understand and keep track of the evolution of the code, improving the understandability of the code and facilitating the debugging process. 
The AI models have shown promising results to generate both description~\cite{Liu2019,jung2021commitbert,nie2021coregen} and its title~\cite{Irsan2022}.
Recent studies also have demonstrated the capability of LLMs, e.g., ChatGPT and GitHub Copiliot~\cite{Xiao2024} to perform these summarization tasks.

\paragraph*{Automated Code Review}
Automated Code review refers to the process of automatically analyzing source code and providing feedback to adhere to coding standards and best practices.
The focus can cover various aspects of code quality such as code style, formatting, performance, security, and maintainability.
Automated code review can help developers catch issues early in the development process, improve code consistency across projects, and ensure that code meets quality standards. 
Several recent works have shown that various sub-tasks of code review can be automated. 
This includes estimating the quality of code~\cite{DBLP:conf/sigsoft/LiLGDJJMGSFS22}, suggesting code refinement~\cite{Thongtanunam2022,Tufano2022}, generating review comment~\cite{DBLP:conf/sigsoft/LiLGDJJMGSFS22,Li2022,Tufano2022}, and suggesting review comment resolution~\cite{Frommgen2024,DBLP:conf/sigsoft/LiLGDJJMGSFS22,Tufano2021,Tufano2022}.
While most of the AI models for code reviews were trained with code review datasets, ChatGPT also has recently shown promising results for performing code review tasks~\cite{Guo2024,Tufano2023}.

\paragraph*{Summary}
As discussed, research has demonstrated that LLMs and AI models can assist developers in enhancing the maintenance and evolution of their human-written code.
This paves a new direction for further improving automatic programming techniques to generate code that meets the non-functional quality for maintenance and evolution. For instance, employing summarization techniques to automatically describe the behavior or the purposes of the generated code to aid developer comprehension. Furthermore, automating code review techniques can be beneficial in assessing the quality of the generated code.

\section{Enhancements of Auto-generated Coding}


 LLMs are not just coding assistants; they have evolved to become versatile partners in the software development process. However, despite the significant strides made by current LLMs, the journey toward their full integration into real-world software development is still lined with challenges. The ``last mile'' of enhancement is crucial for the seamless application of LLMs in practical programming endeavors. As we explore the future progression of LLMs in automatic programming, our roadmap encompasses several pivotal areas of development, aimed at unleashing these intelligent systems to their utmost potential.

\paragraph*{Multi-modal coding} The first area is multi-modal coding. Currently, code LLMs are limited to handling textual data. However, it is crucial to recognize that developers often work with multi-modal data during the development process. For example, The generation of software UI from images and videos requires LLMs to analyze visual elements, understand their context, and transform them into code. This capability would empower developers to streamline the UI design process by simply providing visual examples or prototypes and query LLMs to generate the corresponding code automatically. In addition to UI generation, multi-modal coding has broader implications for software development. Consider the use of figures, tables, and flowcharts in the requirement and design phases. LLMs equipped with multi-modal capabilities could analyze these visual representations and convert them into code snippets automatically. Moreover, multi-modal coding would enhance the interaction between developers and AI models. Developers could foster better collaboration with AI models by communicating their ideas and requirements using multi-modality information. This would enable a more natural and intuitive interaction. The integration of multi-modal coding in LLMs has the potential to revolutionize the software development process. By bridging the gap between visual design and code implementation, LLMs can significantly improve productivity, code quality, and the overall user experience.

\paragraph*{Domains} Secondly, we focus on the empowerment of large-scale domain-specific software. In practical development and maintenance processes, developers often encounter large-scale projects that require diverse domain knowledge. This necessitates customizing LLMs to effectively manage and navigate the complexities of these projects. Software developers frequently grapple with intricate software development challenges within specific business and technology domains, such as e-commerce and automotive. Generating code for such software demands that AI models comprehend various domain-specific concepts. By effectively incorporating specialized domain knowledge into LLMs, these models can provide developers with even more accurate and relevant support. However, handling large-scale projects presents an additional challenge. The current limitations in context length make it difficult for LLMs to process code within large-scale software projects. Even with longer context, comprehending and locating essential information within such extensive code remains a challenge~\cite{liu2024lost}. Overcoming these limitations is crucial to maximize the potential of LLMs and ensure they can effectively meet the demands of complex projects in practical use.

\paragraph*{Knowledge Update}
The third strategic area involves the knowledge repair and updating capabilities of LLMs. LLMs are renowned for their large model size. For instance, GPT-3 has 175 million parameters, requiring an investment of approximately 4.6 million dollars in training and emitting 552 tons of carbon dioxide, equivalent to the emissions of 123 gasoline-powered passenger vehicles driven for one year~\cite{DBLP:journals/corr/abs-2104-10350}. Nonetheless, the evolution of APIs and programming introduces a continuous stream of new knowledge, essential for providing up-to-date services to developers. Moreover, during the maintenance process of code repositories, pre-trained models can unintentionally encounter incorrect information that was previously undiscovered. This can occur when the training code contains undetected buggy codes, leading the model to learn and potentially incorporate inaccurate knowledge. Consequently, the quality of the generated code may be also degraded. Therefore, effectively editing the knowledge of large generative AI models, rather than resorting to periodic retraining of these models from scratch, represents a significant and relatively unexplored research area.

\paragraph*{Reliability and Program Repair}

The fourth focus area is the quality and reliability assurance for the content generated by LLMs. Despite language models' proficiency in code generation, their inherent black-box nature raises concerns about the correctness of the generated code. The increasing reliance on automated programming underscores the need for output that meets the highest standards of quality. This pursuit is not limited to the accuracy of the code; it extends to ensuring the code's maintainability, performance, and scalability. Therefore, enhancing the reliability of the code and creating automated methods to assess and verify the quality of the LLM-generated code is of vital importance.

Overall, we would like to make the following two projections: 
\begin{itemize}
\item     There is a place for last mile repair of auto-generated code using automated program repair techniques \cite{Fan23}.
\item  There remains the enticing possibility of the last mile repair of auto-generated code providing evidence of correctness of the ``improved'' code. This evidence of correctness may be in the form of a test suite which is generated as a by-product of the automated program repair process. We note that such test suites generated in the literature as a by-product of automated program repair have been studied \cite{concolic21}.
\end{itemize}

\paragraph*{Security}
The fifth area of improvement involves security alignment, which is a crucial component of the ``last mile'' of enhancement. It is imperative to address trust issues that arise due to the generation of sensitive or insecure content and potential privacy risks. One of the primary concerns is the generation of insecure code that contains vulnerabilities, which could lead to the crash of software. Additionally, privacy protection is of utmost importance when utilizing code generated by LLMs. The vast amount of data and information processed by these models raises concerns about the handling and storage of sensitive user data. Users' privacy must be safeguarded to ensure that LLMs do not inadvertently leak or misuse personal information. The prevention of harmful content is not only a technical challenge, but also an ethical issue, ensuring that these powerful tools contribute positively to the software development community. Therefore, the objective of security alignment is to design LLMs in a way that avoids generating potentially harmful or insecure content, thus improving the trustworthiness of LLMs and facilitating their widespread adoption. We note that both the reliability and security of the LLM generated code are fundamental to the trustworthiness of LLM output examined in Section \ref{sec:trust}.

\paragraph*{Datasets}
Finally, as different LLMs are trained using different benchmarks, the preparation of high-quality and multidimensional data sets is the key to a fair evaluation of the code generated automatically. In general, current benchmarks focus mainly on highlighting the limitations of the code generated by LLMs. We foresee that the next milestone of LLMs is to generate more complex code and to resolve more complex GitHub issues. With the evolution and increasing capability of LLMs, we foresee newer benchmarks that focus on newer capabilities (e.g., generating code from multi-modal inputs) or newer domains (e.g., autonomous devices). 

\smallskip

In conclusion, the roadmap for enhancing Large Language Models in automated programming is both ambitious and essential. By addressing these six key perspectives, we can anticipate a future where LLMs are not only more capable but also more aligned with the nuanced and evolving needs of software development.


\section{Datasets}


\begin{table*}[h]
\caption{Datasets for Code Generation}
\label{tab:dataset}
\resizebox{\textwidth}{!}{
\begin{tabular}{|l|p{2cm}|p{2cm}|p{2cm}|p{2cm}|p{2cm}|p{3cm}|}
\hline
\small
Benchmark &
  Natural Languages &
  Programming Languages &
  Supported Tasks &
  Size &
  Test case &
Unique Features \\ \hline
APPS~\cite{hendrycks2021measuring} &
  English &
  Python &
  Text-code &
  10,000 problems &
  130,000 total test cases &
  One of the earlier dataset with crowd-sourced   questions for program synthesis \\ \hline
HumanEval~\cite{chen2021evaluating} &
  English &
  Python &
  Text-code &
  164 problems &
  Average 7.7 tests per problem &
  Handwritten problems to evaluate   functional correctness and measure problem-solving capabilities \\ \hline
MBPP~\cite{austin2021program} &
  English &
  Python &
  Text-code &
  974 python functions &
  3 test cases for each problem &
  Measure the ability of these models to   synthesize short Python programs \\ \hline
CONCODE~\cite{iyer2018mapping} &
  English &
  Java &
  Text-code &
  100,000   (classes, NL, code) tuples &
  No test &
  Classes from diverse domains\\ \hline
PandasEval, NumpyEval~\cite{zan2022cert} &
  English &
  Python &
  Text-code &
  101 programming problems &
  20 tests for each problem &
  Library-oriented code generation \\ \hline
MCoNaLa~\cite{wang2022mconala} &
  Spanish, Japanese, and Russian &
  Python &
  Text-code & 
  896 NL-Code pairs & No test
   &
  Support several natural languages beyond English \\ \hline
LLMDefects~\cite{fan2023automated} &
  English &
  Java &
  Text\&code-code &
  113 programming tasks from recent contests, 335 incorrect   solutions &
  1-3 public tests for each problem &
  Contains mistakes in code generated by LLMs.\\ \hline
ClassEval~\cite{du2023classeval} &
  English &
  Python &
  Text-code &
  100 tasks &
  Contain method-level and class-level   tests &
  Class-level code generation \\ \hline
AixBench~\cite{hao2022aixbench} &
  English, Chinese &
  Java &
  Text-code &
  175 samples for automated Test, 161 NL   Task Description &
  Contain hand-crafted tests &
  Contain hand-crafted automated test   cases \\ \hline
  MultiPL-E~\cite{multiple} &
  English &
  19 languages (e.g., Julia, Swift) &
  Text-code &
  161 problems from HumanEval~\cite{chen2021evaluating}, 974 from MBPP~\cite{austin2021program}  &
  Use tests from prior benchmarks~\cite{chen2021evaluating,austin2021program} &
  Extend HumanEval~\cite{chen2021evaluating} and
MBPP~\cite{austin2021program} to 18 languages by translating programs and unit tests\\ \hline
SWE-Bench~\cite{jimenez2023swe} &
  English &
  Python &
  Text\&code-code&
  2294 problems from 12 projects &
  Average 120.8 total tests for each problem &
  Evaluate the ability to resolve real-world GitHub\\ \hline
CodeScope~\cite{yan2023codescope} &
  English &
  43 languages &
  8 tasks&
  200--5,382 samples for each task &
  Contain tests for some tasks &
  Evaluate generated code on difficulty, efficiency,
and length\\ \hline
NoFunEval~\cite{singhal2024nofuneval} &
  English &
  Python, Java, C, JavaScript, Kotlin&
  Text\&code-code, classify correctness &
  47--397 samples for each task &
  No test &
  Evaluate non-functional requirements (latency, security, efficiency)\\ \hline
LiveCodeBench~\cite{jain2024livecodebench} &
  English &
  Python &
  Text\&code-code &
  Collect new problems over time &
  Use tests from programming problems or LLM-generated tests &
  Mitigate contamination issues by crawling new problems\\ \hline
\end{tabular}}
\end{table*}
We perform a literature review of the available datasets for evaluating and studying code generation models. Specifically, we started by doing a preliminary search with keyword ``code generation dataset'' on Google Scholar, we selected all relevant papers and then traced other related work using a backward snowballing approach. We further filter benchmarks that are collection of multiple datasets (e.g., CodeXGLUE~\cite{DBLP:journals/corr/abs-2102-04664}) as their characteristics will be captured by the original datasets in which the collection has been derived from. 

Table~\ref{tab:dataset} shows the existing datasets for code generation. Among existing datasets, HumanEval~\cite{chen2021evaluating} and MBPP~\cite{austin2021program} are one of the earliest benchmarks in which newer benchmarks have been derived from (e.g., MultiPL-E~\cite{multiple} extends HumanEval and MBPP by supporting more programming languages). The HumanEval benchmark contains manually written problems in which the OpenAI Codex model has been evaluated. Recently, SWE-Bench~\cite{jimenez2023swe} was proposed to evaluate whether LLMs can be used to automatically resolve real-world GitHub issues. Based on the reported findings of studies conducted in these datasets, we notice that most of these datasets shows \emph{the limitations of existing LLMs in solving code-related tasks}, highlighting the needs for revolutionary techniques that can further improve these LLMs. For example, the evaluation on SWE-Bench~\cite{jimenez2023swe} shows that their fine-tuned model SWE-Llama can resolve only the simplest GitHub issues. We note that the SWE-bench is gaining attention from practitioners who attempt to automate software engineering beyond a single prompt. A very recent startup effort called Devin \cite{devin} reports reasonable efficacy on SWEbench in autonomously fixing GitHub issues (bug fixes and feature additions). The open-source agent AutoCodeRover \cite{acr} reports higher efficacy than Devin, by considering code structure in localization and fixing.

In general, we observe that most existing datasets require several software artifacts: (1) natural language descriptions (mostly English-centric), (2) code written in a commonly-used programming languages (mostly focus on Python and Java), (3) test cases to verify the correctness of the generated programs.
As shown in the ``Supported Tasks'' column, most existing datasets support text-to-code tasks (code generation from natural language description). 

Although some benchmarks~\cite{iyer2018mapping,wang2022mconala} use textual similarity between the ground truth program and the generated program for validating the correctness, the ``Test case'' column of Table~\ref{tab:dataset} shows that most benchmarks rely on test cases for validating the correctness of generated programs. 
 These test cases are either (1) hand-crafted or (2) translated from other programming languages. Overall, we observe that \emph{a test-driven approach has been widely used for validating the correctness of the generated programs}. This indicates the importance of improving the quality of the test suites used for guiding the code generation. 
Based on the column ``Unique Features'', we observe that recent data sets typically add a new dimension to study the effectiveness of LLMs under a specific condition (e.g. supporting diverse sets of natural languages~\cite{wang2022mconala}, studying defects in automatically generated code~\cite{fan2023automated}). 
Investigating the diverse perspectives of code generation models helps to point out the limitations and the potential bias of the LLMs. 

As most LLMs are trained using programs from open-source repositories, one of the key challenges of a dataset for evaluating the effectiveness of code generation for LLMs is the data leakage problem (e.g., overfitting the training data). Existing datasets usually solve this by using (1) handwritten~\cite{chen2021evaluating} or crowd-sourced problems~\cite{hendrycks2021measuring}, or (2) recently published problems~\cite{fan2023automated}. Recently, LiveCodeBench~\cite{jain2024livecodebench} has been proposed to mitigate data leakage (known as the contamination problem in the article) by continuously crawling new problems from programming contest platforms (\textsc{LeetCode}, \textsc{AtCoder}, and \textsc{CodeForces}). 


\section{Future: Programming environment of 2030-35 and beyond}


In the programming environment of 2030-35 where LLM-based auto-programming techniques have reached certain level of maturity, programmers may need to switch to different roles to fully utilize the power of auto-programming.

\begin{figure}
        \centering

\begin{tabular}{cc}
\begin{tabular}{c}
        \includegraphics[width =0.35\textwidth]{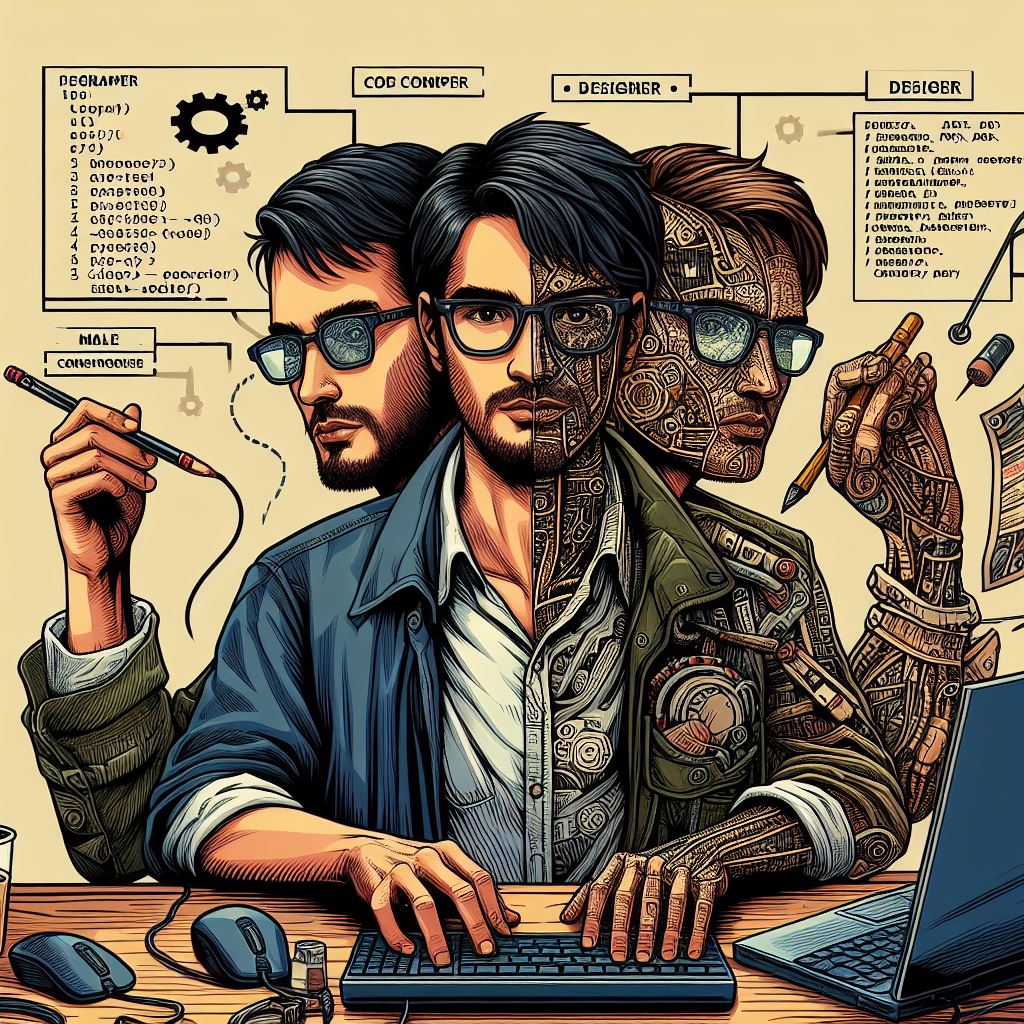}
\end{tabular} &
\begin{tabular}{c}
\includegraphics[width=0.35\textwidth]{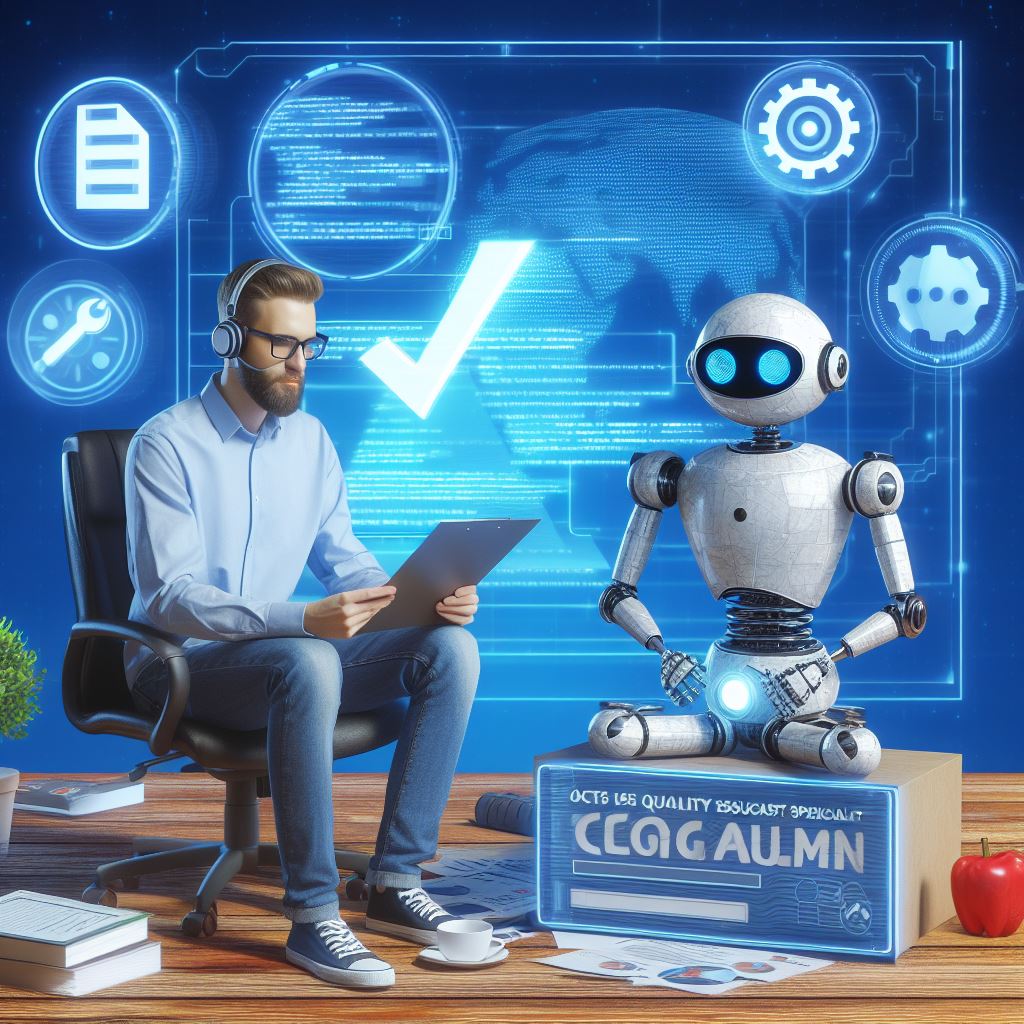}
\end{tabular}\\
(a) & (b) \\
\end{tabular}
        \caption{Evolution of programmer roles captured by DALL-E;
        programmer role (a) as code composer and designer instead of code writer, 
        (b) as quality assurance specialist instead of code writer.
        }\label{combined}
    \end{figure}

\paragraph*{Programmer as code composer and designer instead of code writer.} With the advancement of LLM-based auto-programming, many software maintenance tasks that require code writing can be automatically solved by invoking the appropriate LLMs. Figure~\ref{combined}(a) shows an AI-generated picture by Image Creator that uses DALL·E where programmers acts as code composer and designer instead of code writer. Instead of playing the traditional role of a programmer who meticulously writes code for solving different tasks, they can focus on tasks that require high-level understanding of the requirements (e.g., designing the overall structure of the program and tentative algorithms), allowing automated tools to select the most effective model for the relevant maintenance tasks in which relevant code will be automatically generated. As current techniques mainly focus on specializing LLMs for a specific downstream task of auto-coding (e.g., program repair, log statement, test generation) to improve the effectiveness for the given task, a future programming environment will intelligently predict and select the appropriate model to invoke based the context of the downstream task. There are two scenarios in which LLM-based auto-programming can change the future programming environment: (1) in an Integrated Development Environment (IDE) setting, (2) in a continuous integration (CI) workflow. For example, in the IDE setting that requires instant feedback from the auto-coding tool for efficient interaction, future techniques can design a lightweight tool that can automatically complete and suggest relevant code snippets based on the (1) current surrounding code and (2) the list of available tasks (e.g., suggest adding a JUnit test for the newly written Java method or adding a log statement before a graceful exit of a program).  Meanwhile, in a CI workflow, certain event that represents abnormal behavior of a software system (e.g., test failures, build failures) can automatically trigger the need for a software maintenance task (e.g., the need for a repair can be triggered after a test failure). In this scenario, more sophisticated techniques can be used to further confirm the validity of the trigger (e.g., to distinguish between test failure or flaky test). These techniques include program analysis techniques (such as symbolic execution), test generation techniques (based on code changes within a commit), and log analysis (program monitoring).

\begin{figure}
        \centering
    \includegraphics[width=0.65\textwidth]{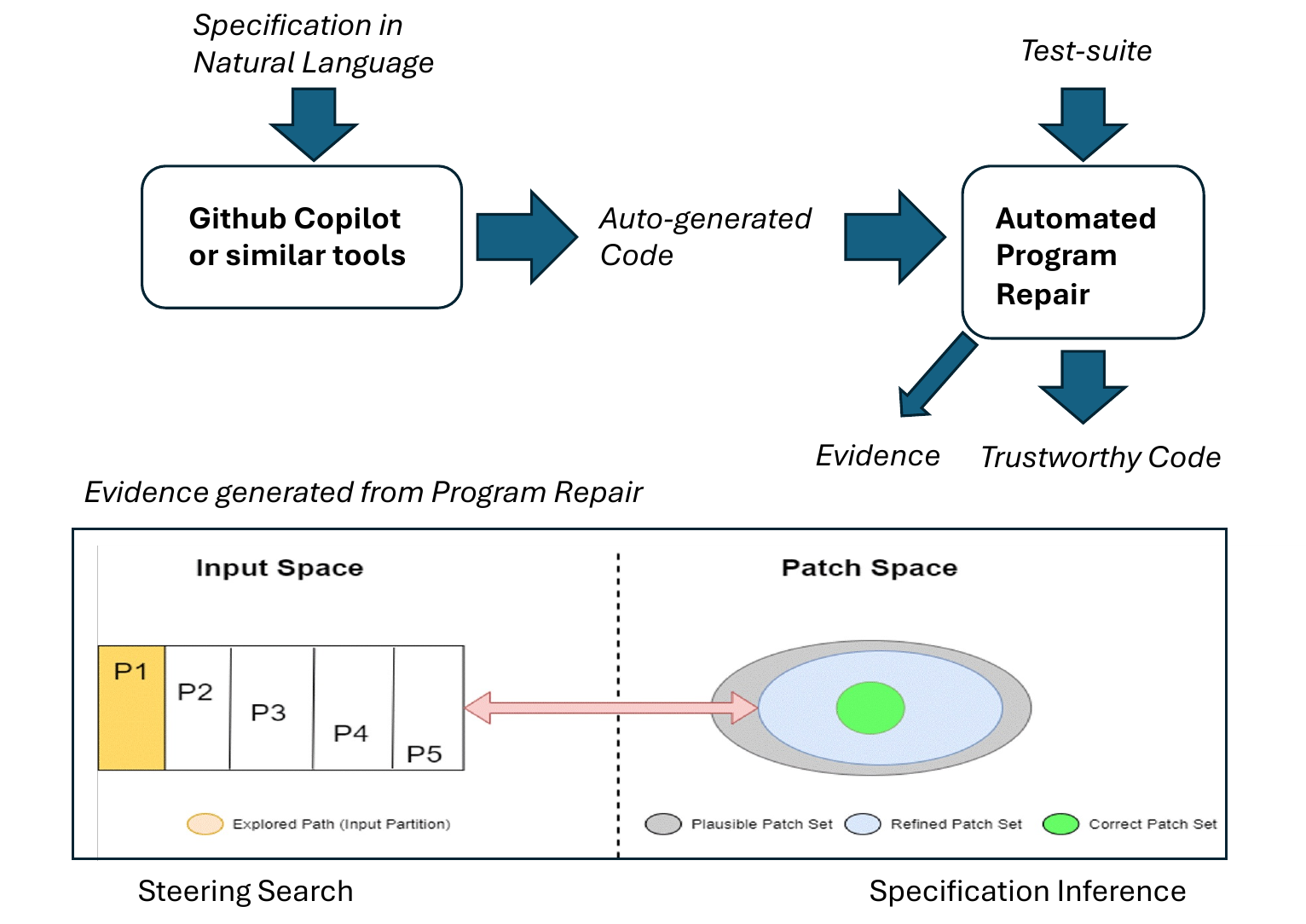}
        \vspace*{-0.05in}
        \caption{Automated repair of auto-generated code}\label{lastmile}
    \end{figure}
\vspace*{-0.05in}
\paragraph*{Programmer as quality assurance specialist.} Although many tasks can be automated, we foresee that concerns regarding the quality of the autogenerated code still remain. As some of the autogenerated code can be misaligned with the intention of the programmer, the programmer will need to play the role of a quality assurance specialist and spend more time in checking the validity of the generated code. Apart from using traditional testing and static analysis tools, more specialized automated program repair techniques can be designed (e.g., by referring to prior study~\cite{fan2023automated} that investigated the mistakes of auto-generated code) to reduce the time and effort involving in checking the quality of auto-generated code.  Figure~\ref{combined}(b) shows an AI-generated picture for programmer main role as quality assurance specialist.
Figure \ref{lastmile} shows a schematic that concretizes this last-mile improvement of autogenerated code $P$ produced from natural language description using a tool like Copilot. The autogenerated code may be subject to program repair guided by a given test suite $T$. However, the process of repair inspects or examines (either explicitly or implicitly) a domain of program edits --- trying to shrink the space of candidate edits which are suitable for improving program $P$. In this way, a repaired program $P'$ is generated from $P$. In the process of examining the domain of program edits (and presumably ruling out a lot of candidate edits), the program repair process generates many additional tests $T'$ over and above the test suite $T$ which was used to guide the program repair process. The oracle (or expected behavior) of these additional tests $T'$ can be obtained via some processing of the natural language description from which $P$ is derived.   The additional test inputs $T'$ (along with their oracles) can then serve as evidence of "correctness" of the repaired program $P'$.  We envision that code-generators of the future will not be only LLMs, but  LLM agents augmented with program analysis / repair capabilities. These augmented code generators may then try to commit code like human programmers, while submitting evidence in the form of generated tests $T'$ as evidence of correctness of (LLM-induced) code commits.  

\vspace*{-0.05in}
\paragraph*{Programmer-assisted safe auto-coding} 
We envision that Large Language Model (LLM) generated code may be integrated into legacy code-bases of existing software projects. This could be in the form of libraries performing specific tasks, where the library code is generated with LLM. For safe integration of such LLM generated code in human-written software projects, one may need {\em sanitizer code} (e.g. \cite{asan}) so that the LLM generated code can be used by the bigger software project safely. Until we reach the stage of completely automating the generation of entire software projects, there may be a need to study (a) automated repair or improvement of LLM generated code, or (b) executing LLM generated code in a {\em contained} manner so that it can cause limited harm to the rest of the software system, or (c) generate verified LLM code whenever appropriate formal specifications are available. One first step towards verified LLM code can be to generate code in a programming language supporting verification ({\em e.g.}, see \cite{Misu24}). We could also generate both programs and proofs (about the program satisfying some formal properties) from LLMs. Such formal properties may be obtained from natural language, in which there is some work \cite{Mauro}. Automated generation of proofs from LLMs has also been recently studied \cite{Baldur}. All of these works provide impetus in moving towards higher assurance code from LLMs. 

\vspace*{-0.05in}
\paragraph*{Autonomous Program Improvement} We view the approach of program repair on unsafe code generated by LLMs to be a more {\em flexible} approach for automatically generating safe code (as it is based on code transformations), as compared to tuning or restriction of LLMs to generate safe outputs. Moving forward, researchers can examine this line of work, in addition to significant short-term efforts in prompt engineering and LLM tuning. Repair of the automatically generated code based on tests gives us more flexibility, partly because we can also choose the tests we use to guide the program repair. The repair as well as other tasks like feature addition can be achieved autonomously by LLM agents which are aware of the structure of the code. The recent work on AutoCodeRover \cite{acr} is an example work in this direction. In the near future, the focus will be to improve the efficacy of these agents. The combination of auto-coding from  natural language and autonomous software improvement using LLMs, is an enticing possibility which can be achieved by 2030. This would shift the role of a future software engineer towards achieving {\em assured autonomy} by focusing on {\em trust} of the autonomous artifacts, instead of engineering software systems at {\em scale}. The scale of software systems is likely to be achieved automatically in future, thus shifting the attention to \underline{\em trust}.


\paragraph*{Looking even further} Autonomous improvement of automatically generated code need not be restricted to application level programming. We could examine the feasibility of automatically repairing probabilistic programs, which are generated automatically. Probabilistic  programming succinctly expresses statistical inference tasks on probabilistic models \cite{fose14}, and are supported in the back-end by machine learning frameworks like Pytorch \cite{pytorch}. Recently symbolic execution of probabilistic programs was proposed \cite{Voogd23} which raises the possibility of semantics-aware probabilistic program repair, after an initial LLM guided automated generation of  probabilistic program snippets. This line of work could help us progress towards automated self improvement of learning tasks,  a speculative direction of future research.





\section*{Acknowledgments}
This work is partially supported by a Singapore Ministry of Education( MoE) Tier3 grant  MOE-MOET32021-0001. 
 The authors thank Prem Devanbu for his valuable comments about the article. The corresponding author Abhik Roychoudhury would like to thank Xiang Gao and Martin Mirchev for contributing some example programs to illustrate the issues with AI based coding.

\bibliographystyle{ACM-Reference-Format}
\bibliography{references, ref-baishakhi, references-added}

\end{document}